\documentstyle[twoside,fleqn,jltp,epsf]{article}

\def \br{{\bf r}}
\def \brp{{\bf r'}}
\newcommand{\omr}{{\bf\Omega_\br}}
\newcommand{\omrp}{{\bf\Omega_\brp}}
\def \virg{\;\;,}
\def \point{\;\;.}
\newcommand{\bbox}[1]{\mbox{\boldmath $#1$}}
\newcommand{\fra}[2]{\mbox{$\frac{#1}{#2}$}}
\newcommand{\iob}{\int_0^\beta d \tau}
\newcommand{\ve}[1]{\makebox{\boldmath{$#1$}}}
\def \omn#1{\bbox{\Omega}_{#1}}

\title{Functional Integrals for Correlated Fermions}

\author{H.J. Schulz\address{Laboratoire de Physique des Solides,
Universit\'e Paris--Sud, 91405 Orsay, France}}

\runninghead{H.J. Schulz}{Functional Integrals for Correlated Fermions}

\begin{document}

\input amssym.def
\input amssym.tex

\epsfverbosetrue

\begin{abstract}
Functional integral methods provide a way to define mean--field theories and
to systematically improve them. For the Hubbard model and similar
strong--correlation problems, methods based in particular on the
Hubbard--Stratonovich transformation have however been plagued by
difficulties to formulate the problem in a spin--rotation invariant
way. Here a formalism circumventing this problem by using a space-- and
time--dependent spin reference axis is discussed. This formulation is
then
used to suggest a possible alternative to Nagaoka ferromagnetism in the
strongly correlated Hubbard model in the vicinity of half--filling. Finally,
some aspects of single--particle spectra in a simplified model for a
short--range ordered antiferromagnet are discussed.

PACS numbers: 71.28.+d, 75.10.Lp, 75.30.Kz
\end{abstract}

\maketitle

\section{INTRODUCTION}
It has been pointed out soon after the discovery of high--temperature
superconductivity in the cuprates that electron--electron correlations are
likely to play the crucial role in determining the physics of these
compounds.  \cite{anderson_hgtc_hubbard} This observation is based on the
fact that compounds with a nominally half--filled band like $\rm La_2CuO_4$
are antiferromagnetic insulators rather than metals (as would be expected
from band theory). This of course is explained by the well--known
Mott--Hubbard effect, implying important correlation effects. Rather small
amounts of doping away from half--filling than lead to an unusual metallic
state which becomes superconducting at high temperatures.

As pointed out by Anderson,\cite{anderson_hgtc_hubbard} the (strongly
correlated) Hubbard model can be expected to reproduced many of the
essential features: at half--filling the model can be mapped onto the
antiferromagnetic Heisenberg model and therefore gives rise to the
antiferromagnetic insulating state, and it is equally clear that doping will
eventually lead to a conducting state. How exactly this happens is however
not clear at present: for weak correlation doping initially gives rise to an
incommensurate antiferromagnetic (or spin density wave) {\em insulator}, and
only upon further doping is a conducting state
reached.\cite{schulz_gf_incomm} For stronger correlation, the situation may
well be different. In particular, the spiral state proposed by Shraiman and
Siggia\cite{shraiman_tj_spirale_bis} is conducting already at the smallest
doping level, even though long--range magnetic order may persist. Moreover,
neither the anomalous properties of the normal state in the conducting
regime nor the origin of superconductivity are currently well--understood in
the framework of the Hubbard model or of its possible generalizations.

Whatever the critical value of doping $\delta_c$ at which magnetic order
disappears, one can expect that for $\delta \gtrsim \delta_c$
short--range antiferromagnetic order with a more or less important
correlation length $\xi$ still exists. It is then important to
understand the electronic properties in this regime, in particular in
view of understanding the unusual properties of the metallic state of
the cuprates. Moreover, the ordered state at half--filling can be
understood in quite some detail (even for strong correlation) from
Hartree--Fock theory supplemented by spin--wave corrections. At least for
large (but finite) $\xi$ it therefore seems reasonable to start from a
description
of the ordered state and then, in a second step,  to add the effects of the
fact that the order is only short ranged. Ideally, this might even help to
understand the situation of very short $\xi$. In fact for large $U$ (and
hole doping) a site is either empty and thus non--magnetic or singly
occupied and thus represents, on some short time scale, a magnetic moment.
The physics of strong correlation thus is closely related to that of more or
less short ranged magnetic order. In fact, the dynamics of the strongly
correlated fermions can be seen as that of fermions moving in a space-- and
time--dependent spin ``background'' which however is generated by the
fermions themselves and therefore has to be determined self--consistently.

In the following section a generalization of the standard
Hubbard--Stratonovich functional integral
formalism\cite{hubbard_trans,strato_trans} will be described that explicitly
exhibits the local spin dynamics and therefore can be expected to make a
formally satisfying treatment of the situation discussed above
possible.\cite{schulz_su2,schulz_sanseb} Some simple applications will then
be discussed in the subsequent sections.

\section{FORMALISM}
The standard Hubbard Hamiltonian has the form
\begin{equation}
H = -t \sum_{\langle \br\brp \rangle} (a_{\br\sigma}^\dagger
a^{\phantom{\dagger}}_{\brp\sigma} + h.c.)  + U \sum_\br n_{\br\uparrow}
n_{\br\downarrow} \virg
\end{equation}
where $a_{\br\sigma}^\dagger$ creates a fermion at site $\br$ with spin
projection $\sigma$, $t$ is the nearest neighbor hopping integral, $U$ is
the onsite interaction, and $\langle
\br\brp \rangle$ indicates summation over nearest--neighbor bonds, each bond
being counted once. Introducing a spinor notation via
\begin{equation}
\Psi_{\br} = \left(
\begin{array}{c}
\psi_{\br \uparrow} \\
\psi_{\br \downarrow}
\end{array}  \right)
\end{equation}
the partition function can be written as a functional integral over
Grassmann variables:
\begin{equation}
Z = \int {\mathcal D} \Psi e^{-S_0-S_{int}} \point
\end{equation}
The free and interaction parts of the action are respectively
\begin{equation}
S_0 = \int_0^\beta d\tau \left\{\sum_{\br} \Psi_{\br}^*(\partial_\tau - \mu)
\Psi^{\phantom{*}}_{\br} - t \sum_{\langle \br\brp \rangle }(\Psi^*_{\br}
\Psi^{\phantom{*}}_{\brp} + c.c.)  \right\} \virg
\end{equation}
and
\begin{equation}
S_{int} = U \int_0^\beta d\tau \sum_{\br}
\psi^*_{\br \uparrow}\psi^*_{\br \downarrow}
\psi^{\phantom{*}}_{\br \downarrow}\psi^{\phantom{*}}_{\br \uparrow} \point
\end{equation}
In the following, the interaction term will be treated using a
Hubbard--Stratonovich decomposition.\cite{hubbard_trans,strato_trans}

For repulsive interactions ($U>0$) the appropriate decomposition is
\begin{equation}
e^{-\varepsilon U \psi^*_{\uparrow}\psi^*_{ \downarrow}
\psi^{\phantom{*}}_{\downarrow}\psi^{\phantom{*}}_{\uparrow}}
=  \frac{\varepsilon}{\pi U}
\int d\Delta_c d\Delta_s \exp \left[-\frac{\varepsilon}{U}(\Delta_c^2
+ \Delta_s^2) + i\varepsilon\Delta_cn + \varepsilon\Delta_s\sigma_z
\right] \point
\label{hs2}
\end{equation}
Here $\Delta_{c,s}$ are real variables, and $n = \psi^*_\uparrow
\psi^{\phantom{*}}_\uparrow + \psi^*_\downarrow
\psi^{\phantom{*}}_\downarrow$, $\sigma_z = \psi^*_\uparrow
\psi^{\phantom{*}}_\uparrow - \psi^*_\downarrow
\psi^{\phantom{*}}_\downarrow$. One inserts this
at each point in space and time and thus obtains a functional integral over
charge and spin fields $\Delta_{c,s}(\ve{r},\tau)$, coupled bilinearly
to the
fermions. A saddle point approximation reproduces the Hartree--Fock results,
and in particular at half--filling one finds an antiferromagnetic (or
spin--density wave in another terminology) ground state. The unpleasant
feature of this way of proceeding is that both $\Delta_c$ and $\Delta_s$ are
scalar fields, and one therefore cannot construct easily the effective
action for the low--energy excitations of the antiferromagnetic state which
are spin--waves, the existence of which is of course closely related to the
vectorial character of the order parameter.

Alternatively, one might use a Hubbard--Stratonovich decomposition using a
vector auxiliary field. One then however does not even obtain the
Hartree--Fock solution as a saddle point. A number of other, equally
unsatisfactory decompositions have been discussed in the
literature.\cite{gomes_lederer} In order to obtain a spin--rotation
invariant effective action for fluctuations around the Hartree--Fock
solution, I notice\cite{schulz_su2,weng_su2} that in writing down the
Hamiltonian the choice of the spin quantization axis is a priori arbitrary
at each lattice site, and in a functional integral formulation can also vary
in time. I then leave the quantization axis $\omr(\tau)$ arbitrary and
integrate over all possible $\omr(\tau)$, with the appropriate invariant and
normalized integration measure at each point in space and time.  In
practice, this is achieved by introducing $SU(2)$ rotation matrices
$R_{\br}(\tau)$ at each point of space and time which satisfy
$R^{\phantom{+}}_{\br}(\tau) \sigma_z R^+_{\br}(\tau) = \omr(\tau) \cdot
\bbox{\sigma}$.  A convenient choice is
\begin{equation}
\label{rm}
R(\bbox{\Omega}) = \left(
\begin{array}{cc}
\cos (\fra12 \theta) & -e^{-i \varphi} \sin(\fra12 \theta) \\
e^{i \varphi} \sin(\fra12 \theta) &
\cos (\fra12 \theta)
\end{array}
\right) \virg
\end{equation}
where $\theta$ and $\varphi$ are the usual polar angles.  I then introduce
identities $R^{\phantom{+}}_{\br}(\tau) R^+_{\br}(\tau) = 1$ at the
appropriate places in the functional integral and integrate over all
configurations $\omr(\tau)$. Finally, new spinor variables are introduced
via
\begin{equation} \label{eq:tra}
\Phi = R^+ \Psi \point
\end{equation}
This means that the $\phi$--particles now have their spin along $\pm
\omr(\tau)$. The Hubbard interaction term is invariant under
this transformation, and now the Hubbard--Stratonovich transformation can be
used in its form (\ref{hs2}) without loosing the spin excitations which are
contained in the functional integral over $\omr(\tau)$. This
also means that a nonzero saddle point value of the spin field $\Delta_s$
does not necessarily imply the existence of magnetic long--range order. For
this to occur, the angular degrees of freedom have also to be ordered.

Given that at least at half--filling Hartree--Fock theory
handles {\em local} correlations rather well even for large $U$,
one can now start by using a saddle point approximation for the scalar fields
$\Delta_{c,s}$.  The partition function then becomes
\begin{equation}
\label{Z}
Z = \int {\mathcal D} \bbox{\Omega} {\; \mathcal D} \Phi \; {\mathcal D} \delta
\; e^{-S_{HF}-S_\Omega-S_\delta}
\point
\end{equation}
Here $S_{HF}$ is the action corresponding to the saddle point,
and $S_\Omega$ represents the coupling
between the angular fluctuations and the fermions. Finally
$S_\delta$
represents the massive fluctuations of $\Delta_c$ and $\Delta_s$ around
their respective saddle point values, and will be neglected in the
following.

For large $U$ an effective action for the spin degrees of freedom and the
doped carriers can be derived, because in fact arbitrary space--time
variations of $\bbox{\Omega}(\bbox{r},\tau)$ can be treated. For simplicity,
one can then start from a ferromagnetic saddle point which is characterized
by lower and upper Hubbard band separated by a gap $U$. One then has
\begin{equation} \label{eq:HF}
S_{HF} = \int_0^\beta d\tau \left\{\sum_{\br} \Phi_{\br}^*(\partial_\tau
- \mu + \frac12 U \sigma_z)
\Phi^{\phantom{*}}_{\br} - t \sum_{\langle \br\brp \rangle }(\Phi^*_{\br}
\Phi^{\phantom{*}}_{\brp} + c.c.)  \right\} \virg
\end{equation}
\begin{equation} \label{eq:som}
S_\Omega = \int_0^\beta d\tau \left\{ \sum_\br \Phi^*_\br R^+_\br
\dot{R}^{\phantom{*}}_\br
\Phi^{\phantom{*}}_\br
-t \sum_{\langle \br \brp \rangle} [\Phi^*_\br (R^+_\br R^{\phantom{+}}_\brp
-1)
\Phi^{\phantom{*}}_\brp
+ c.c.] \right\} \point
\end{equation}
For the case of electron doping, the chemical potential is somewhere in the
upper Hubbard bands, and the lower Hubbard band then can be integrated
out.\cite{schulz_su2} In this way one obtains the effective action for the
local spin orientation and particles in the upper Hubbard band order by
order in $t/U$.  To zeroth order in $t/U$ I find
\begin{eqnarray}   \nonumber
S_{\mbox{\it eff}}^0 & = & \iob \left\{ \sum_\br [ \phi^*_\br
(\partial_\tau - \mu + U)
\phi^{\phantom{*}}_\br - \frac{i}{2} \dot{\varphi}_\br ( 1 - \cos
\vartheta_\br ) (1 -
\phi^*_\br \phi^{\phantom{*}}_\br ) ] \right. \\
\label{eq:s0}
& & \left.  - t \sum_{\langle \br \brp \rangle} [ \alpha (\omr , \omrp )
\phi^*_\br \phi^{\phantom{*}}_\brp + c.c.] \right\} \point
\end{eqnarray}
Here $\phi$ refers to fermions in the upper Hubbard band, the spin index
being omitted, and $\varphi_\br , \vartheta_\br$ are the polar angles of
$\omr$. The coefficients $\alpha ( \omr , \omrp )$ come from the expression
for the product of two $R$ matrices and are given by
\begin{equation}
\label{alpha}
\alpha ( \omr , \omrp )  =  |\alpha | e^{i \chi_{\br \brp}}
= [(1 + \omr \cdot \omrp )/2]^{1/2}
 \exp [ i \hat{A}(\omr , \omrp ,\hat{\ve{z}}) /2 ] \virg
\end{equation}
where $\hat{A}(\omn1  , \omn2 ,\omn3 )$ is the signed
solid angle spanned by the vectors $\omn1 , \omn2 , \omn3 $
\cite{berg_luscher}, and
 $ \hat{\ve{z}} $ is the unit vector along $z$. To next order in $t/U$
one recovers the usual $t^2/U$ antiferromagnetic exchange
term.\cite{schulz_su2}

\section{A POSSIBLE INSTABILITY OF THE NAGAOKA STATE}
In the absence of particles in the upper Hubbard
band, in $S^0_{eff}$ only  the purely imaginary term remains, which is
the Berry phase of an isolated spin $1/2$, i.e., as expected, the
half-filled Hubbard model becomes a collection of independent spins for
$U=\infty$. Introducing more fermions, two effects occur: (i) the
factors $(1  -  \phi^*_\br \phi^{\phantom{*}}_\br )$, previously introduced by
Shankar from semi-phenomenological arguments,\cite{shankar_trous_bis}
cancel the Berry phase term
whenever there is an extra particle on site $\br$, i.e. one is in a spin
singlet whenever two particles occupy the same site.
(ii) the kinetic energy
term plays a role: in particular, going around an elementary
plaquette $(1234)$ the lattice curl of the phases
$\chi_{\br \brp}$ equals $\Phi_{1234} = [ \hat{A}(\omn1, \omn2 ,\omn3 ) +
\hat{A}(\omn3, \omn4 ,\omn1 ) ]/2 $, i.e. there is an effective magnetic
field proportional to the solid angle spanned by $\omn1,...,\omn4$.
$\Phi_{1234}$ is the lattice analogue of the familiar winding number
density of the continuum nonlinear $\sigma $ model
\cite{berg_luscher}.
Note that, while the gauge potential in (\ref{alpha}) depends explicitly
on $\hat{\ve{z}}$ and therefore is not rotational invariant, the
physical fluxes are. For coplanar configurations $\Phi_{1234} = 0$,
i.e. the phases
can be removed by a gauge transformation of the $\phi$'s. One then
straightforwardly sees that the kinetic term is optimized by a
ferromagnetic arrangement of the spins. This is the familiar Nagaoka
phenomenon.\cite{nagaoka}

\begin{figure}[htb]
\centerline{
\epsfysize 7cm
\epsffile{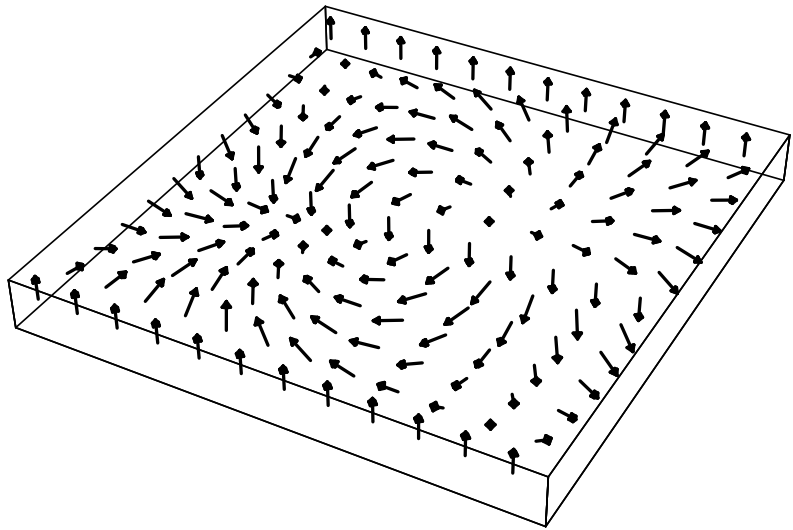}}
\caption{A spin arrangement (``texture'') giving rise to a constant
effective magnetic field acting on the fermions.}
\label{f:1}
\end{figure}

One can now ask the question whether non--coplanar configurations of
$\ve{\Omega}_\br$ with a nonzero winding number density
can lead to an energy lower than the Nagaoka state.
\cite{doucot_ferro_instab} A configuration giving rise  to a
spatially constant
effective field (i.e. a constant lattice curl of $\chi_{\br \brp}$)
seen by the fermions is shown in fig.\ref{f:1}.
At the mean--field level, one assumes the $\omr$ to be static, and
only the first and third terms in eq.(\ref{eq:s0}) contribute. It
is then easy to convince oneself that if this field is such that the
lowest Landau level is completely filled and all the other Landau level
are empty the energies of the ferromagnetic and the textured states are
nearly identical. More precisely, the energy of the textured state is
higher than that of the ferromagnetic state only due to the fact that now
$\omr \cdot \omrp < 1$, i.e. the global scale of the kinetic energy term
is reduced. This leads to an energy difference
\begin{equation} \label{eq:corr1}
\Delta E \approx t n^2 \virg
\end{equation}
in favor of the ferromagnet. Here $n$ is the dopant density, i.e. $n=1$
corresponds to half--filling. However, one should notice that the fully
polarized ferromagnet ($\omr \equiv \hat{\ve{z}}$) is an exact
eigenstate of the Hamiltonian, whereas the textured state is a
mean--field trial state, i.e. its mean--field energy is an upper limit
to its exact energy. In particular, one can calculate the first order
quantum fluctuation corrections \cite{schulz_unpub}: the collective
modes have a spectrum  $\omega(\ve{q}) = v |\ve{q}|$, with a velocity
given by
\begin{equation}\label{eq:v}
v = 2t \sqrt{2\pi n} \point
\end{equation}
This then gives rise to a fluctuation correction to the ground state
energy of order
\begin{equation}\label{eq:corr2}
\Delta E_{\mbox{\it fluc}} \approx -t n^2  \virg
\end{equation}
i.e. of the same order of magnitude but with opposite sign as the
correction due to the band narrowing, eq.(\ref{eq:corr1}). To determine
which of the two corrections is more important, a detailed calculation
taking into account short--distance cutoffs would be necessary. In the
absence of such a calculation one can only point out that a textured
state is a possible candidate for the destabilization of Nagaoka
ferromagnetism.

The existence of a textured state might in fact have interesting
consequences: (i) there is a large class of spin textures that all give
rise to a constant effective magnetic field. It is conceivable that this
large degeneracy gives rise to sufficiently strong quantum fluctuations
to destroy long--range magnetic order. The effective magnetic field seen
by the fermions would however still persist, i.e. the PT symmetry
breaking inherent in a textured state would survive. This then could
lead to a state similar to those proposed in the context of anyon
superconductivity.\cite{laughlin_anyons,lederer_flux} (ii) even if long--range
order exists the true ground state would be a spin singlet, due to the
absence of any spontaneous magnetization. This might explain the strong
variation of the total ground state spin observed in finite--size
studies\cite{riera_young}. Moreover, with {\em twisted boundary
conditions}, the small--doping ground state is in fact found to be
fully polarized in finite--size calculations.\cite{poilblanc_ferro} The twist
necessary to lead to the polarized ground state may well be a remnant of
the spin texture which (hypothetically) is present in the thermodynamic
limit.

\section{SINGLE PARTICLE STATES}
It is clearly of importance to understand the single--particle
properties in a doped and short--range ordered antiferromagnet. Fully
understanding this problem, even based on the simplified action,
eq.(\ref{eq:s0}), and possibly the correction of order
$t/U$,\cite{schulz_su2} is still
a formidable unsolved problem. Nevertheless, some insight can be gained
transforming eqs.(\ref{eq:HF}) and (\ref{eq:som}) back to the original
$\Psi$ variables, via eq.(\ref{eq:tra}). One obtains
\begin{eqnarray}\nonumber
S_{HF} + S_\Omega & = &
 \int_0^\beta d\tau \left\{\sum_{\br} \Psi_{\br}^*(\partial_\tau
- \mu + \frac12 U \omr \cdot \ve{\sigma})
\Psi^{\phantom{*}}_{\br} \right. \\
\label{eq:shfom}
&& \quad \left. - t \sum_{\langle \br\brp \rangle }(\Psi^*_{\br}
\Psi^{\phantom{*}}_{\brp} + c.c.)  \right\} \virg
\end{eqnarray}
\begin{figure}[htbp]
\centerline{
\epsfysize 16cm
\epsffile{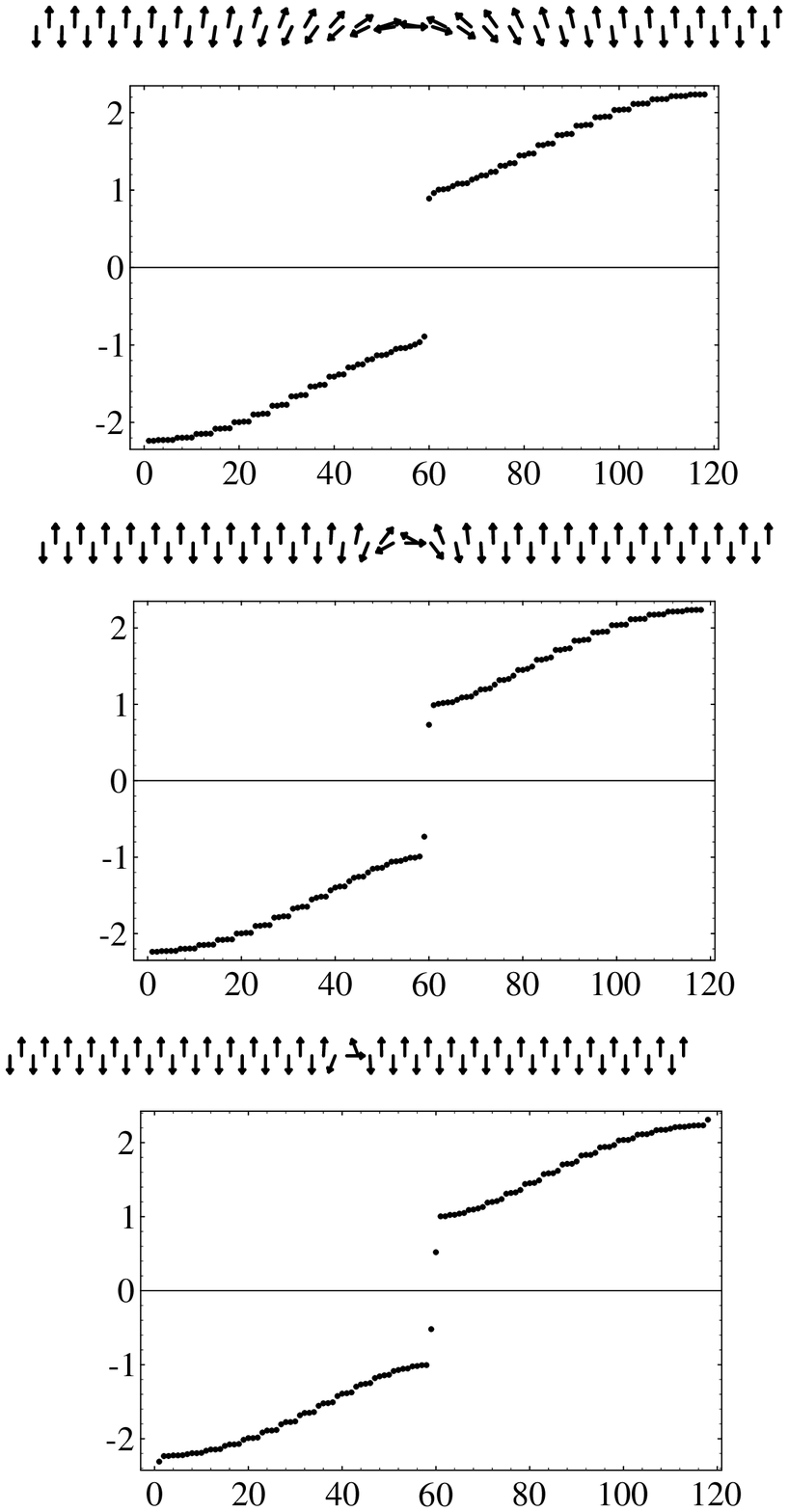}}
\caption[gt]{Single particle states for electrons moving in a magnetic
field of constant magnitude but varying orientation (see
eq.(\ref{eq:shfom})). The spin configurations are shown
together with the corresponding spectrum and represent a defect in
an otherwise antiferromagnetically ordered state. Here in all cases
approximate antiferromagnetic order is preserved even within  the
defect. The width of the defect decreases from top to bottom.
}
\label{f:2}
\end{figure}
\begin{figure}[htbp]
\centerline{
\epsfysize 16cm
\epsffile{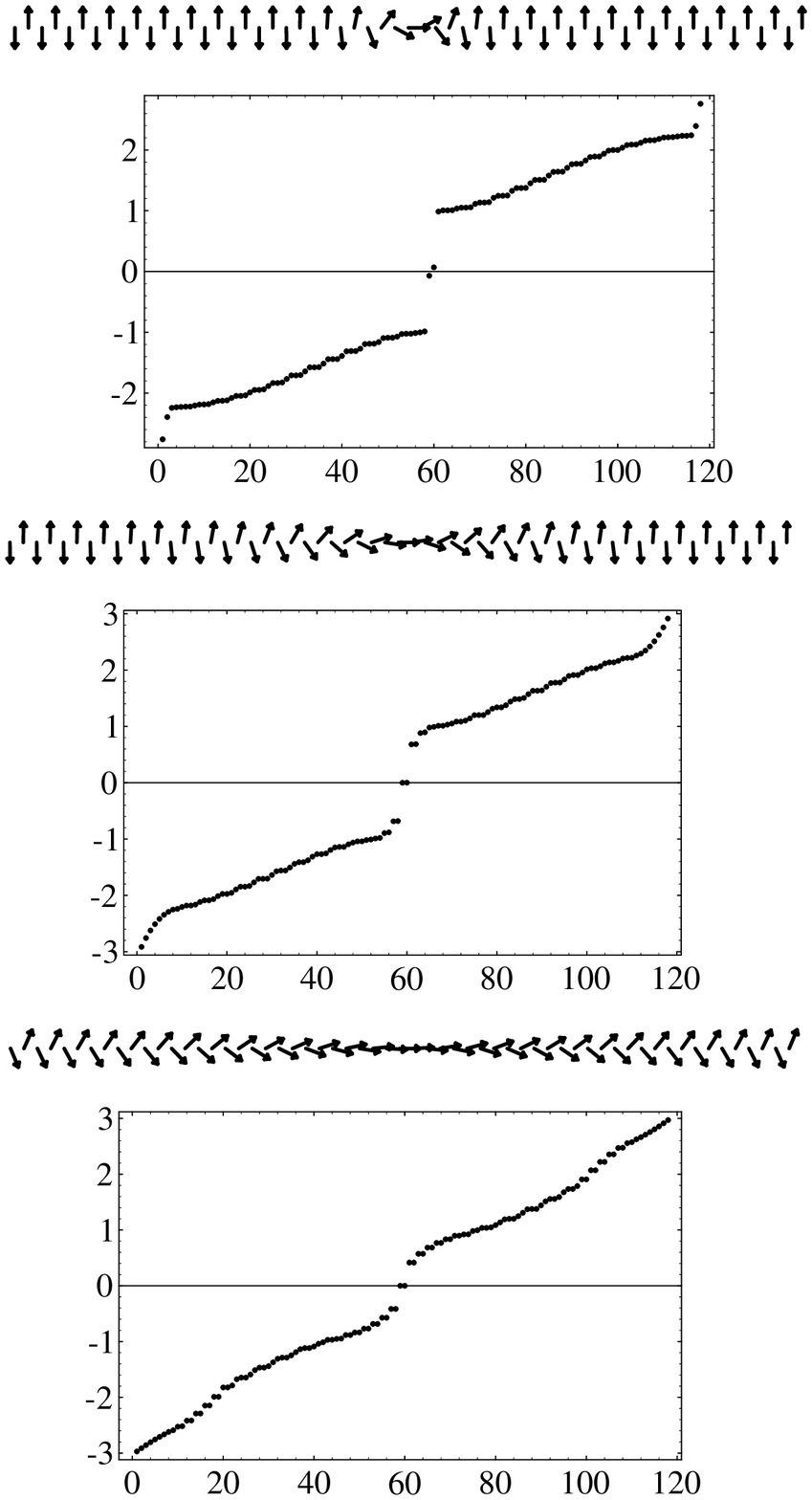}}
\caption{The same as fig.\protect{\ref{f:2}}, but here there is a strong
ferromagnetic component of the magnetization at the defects.}
\label{f:3}
\end{figure}
i.e. effectively a problem of electrons moving in a space and
time--dependent filed of constant magnitude but varying orientation.
A priori all possible time and space dependent configurations
$\omr(\tau)$ should be summed over. However, it seems likely that the
dominant configurations will have some degree of short--range
antiferromagnetic order. At least in one dimension such a state can be
considered
as an antiferromagnet disordered by defects, and I therefore consider
configuration where there is antiferromagnetic order and just one defect
changing the sign of the order parameter. To obtain a calculationally
tractable problem, I further consider static configurations.
In a truely disordered antiferromagnet with some finite correlation
length, one of course has a finite density of defects (provided these
can still be defined), but the present highly simplified calculation may
still give some helpful ideas. In figs.\ref{f:2} and \ref{f:3} I show
some defect configurations together with their electronic spectrum. In
particular, in fig.\ref{f:2}, local antiferromagnetic order is preserved
nearly everywhere. One observes that the separation of states into upper
and lower Hubbard bands remains unless the width of the defect becomes
very small (the lowest panel in the figure). If one the other hand in
the vicinity of the defect there is some form of local ferromagnetic
order (fig.\ref{f:3}), one always has states in the middle of the band.
One can expect that the two rather different types of spectra for local
antiferromagnetic or ferromagnetic alignment persist in higher
dimensions.
The two alternative types of spectra (or spectral densities) thus might
be observed in photoemission
spectroscopy. Moreover, configurations like those of fig.\ref{f:3}
because of their local ferromagnetic order should give rise to an
enhanced spin susceptibility.

In the present calculations the spin structure $\omr$ was assumed to be
static. This is probably a reasonable approximation if the electronic
excitation energy (as measured from the Fermi level) is high compared to a
typical frequency of the spin dynamics. If on the other hand one is at low
excitation energy, i.e. in somewhere in the middle of the spectra of figures
\ref{f:2} and \ref{f:3} if the band is approximately half--filled, the spin
dynamics will be effectively fast compared to the electronic dynamics. One
can then speculate that some form of ``motional narrowing'' will transform
the states close to the center of the spectra into plane--wave like
states,
giving rise to Fermi--liquid or possibly non--Fermi--liquid like properties.
It should also be noted that the spectra shown here are symmetric about the
horizontal axis, in particular there is no ``spectral weight transfer'' from
the upper to the lower Hubbard band. This is due to the neglect of the
charge interaction contained in the $S_\delta$ term in eq.(\ref{Z}).

\section{CONCLUSION}
I have discussed here a way to give a spin--rotation invariant
functional integral formulation of the strong correlation problem.
This formulation makes the role of local (in space and time) magnetic
order particularly apparent. For half--filling, this formulation allows
one to recover the mapping to the antiferromagnetic Heisenberg model for
large $U$, and more generally to obtain the effective spin dynamics for
arbitrary $U$.\cite{schulz_su2,schulz_sanseb} As discussed here
alternatives to Nagaoka ferromagnetism arise at strong coupling, and
properties of single--particle spectra for different types of local
magnetic order can be discussed.

It should be pointed out that the method can be generalized so as to
include the electron--hole symmetry of the Hubbard
model.\cite{schulz_su2,schulz_sanseb} One then obtains a formulation
with a fluctuating reference frame in an $SU(2) \times SU(2)$ space.
This can in particular be used to treat particle--particle and
particle--hole type instabilities on an equal footing in a
Ginzburg--Landau like formulation. For example, one can include
Kanamori--type T--matrix renormalizations into a microscopic formulation
of a Ginzburg--Landau theory of magnetic ordering. Also, the large $U$
limit of the present formulation can be related \cite{schulz_sanseb}
to the slave--fermion Schwinger boson approach.\cite{lee_gauge}

\section*{ACKNOWLEDGMENTS}
I am grateful to T. Giamarchi, A. Ruckenstein, and J.R. Schrieffer for
stimulating discussions.

\end{document}